\documentstyle[12pt]{article}
\newcommand{\beq}{\begin{equation}}
\newcommand{\eeq}{\end{equation}}

\begin{document}

\begin{center}

\vskip3em
{\bf \huge Exact Wavefunctions for a Delta Function
  Bose Gas with Higher Derivatives}

\vskip3em
{\large Eric D. Williams}

\vskip1em

{\it Institute for Solid State Physics, University of Tokyo, 7-22-1
  Roppongi,  Tokyo 106}

{\it email: williams@ginnan.issp.u-tokyo.ac.jp}

\end{center}

\vskip 2cm

\baselineskip7mm

\begin{abstract}
\baselineskip6mm

A quantum mechanical system describing bosons in one space dimension
with a kinetic energy of arbitrary order in derivatives and a delta
function interaction is studied.  Exact wavefunctions for an arbitrary
number of particles and order of derivative are constructed. Also,
equations determining the spectrum of eigenvalues are found.

PACS numbers: 02.30Jr, 05.30Jp

\end{abstract}

\setcounter{equation}{0}

Physical systems whose equations contains derivatives higher than two
arise in a number of contexts.  In classical mechanics, for example,
the Korteweq-deVries equation, which describes shallow water waves,
contains a third order derivative term~\cite{korteweg}. Higher
derivative theories of classical gravity, first introduced by
Weyl~\cite{weyl}, have some attractive cosmological properties, such
as gravity driven expansion~\cite{expand}. A model of quantum gravity
with higher derivative terms has been shown to be
renormalizable~\cite{renorm} and asympotically free~\cite{free}. In
particle physics, the Nambu-Jona-Lasinio~\cite{nambu},
Schwinger~\cite{schwinger}, and Skyrme~\cite{skyrme} models with
higher derivative terms added have been considered. Also, in the
superfield formulation of supersymmetric field theories, higher
derivative terms arise naturally~\cite{ss}.

In this work, a family of higher derivative generalizations of
Schrodinger's equation is studied.  The system describes bosons where
the ``kinetic energy'' is a polynomial of order $n$ in derivatives and
the interaction is a delta function potential.  A particular choice of
polynomial corresponds to the much studied delta function bose gas
(or quantum nonlinear Schrodinger equation)\cite{lieb,many}.  The main
result here is that for an arbitrary polynomial in derivatives and
arbitrary number of particles, exact eigenfunctions for the system may
be constructed.

Formulating the system under study, consider the eigenvalue problem
\beq
      \label{se}         
   H \Psi = E \Psi,
\eeq
where the hamiltonian $H$ is the partial differential operator
\beq
\label{fullhamil}
       H= \sum_{j=1}^{N} \sum_{l=1}^{n} a_l  (-i \partial_{x_j})^l
       + \sum_{j<k} 2c \delta(x_j - x_k),
\eeq
and $\Psi(x_1,...,x_N)$ is a symmetric function of the $x_j$. $N$ is
the number of bosons, $n$ is an arbitrary natural number, the $a_l$
are arbitrary real coefficients, and $ c$ is a real interaction
constant.  Equation (\ref{se}) is a generalization of Schrodinger's
equation describing bosons interacting via a delta function where the
usual 2nd order kinetic energy has been replaced with a polynomial
containing higher order derivatives.

The hamiltonian~(\ref{fullhamil}) is translationally invariant, thus the 
momentum operator
\beq
P=-i \sum_{j=1}^{N} \partial_{x_j}
\eeq
commutes with the hamiltonian~(\ref{fullhamil}). 

If $a_2=1$, and all other $a_l=0$, then (\ref{fullhamil}) is called the
$\delta$-function bose gas, and the eigenfunctions for arbitrary $N$
were first found by Lieb and Liniger~\cite{lieb}.  The method of
solution has come to be known as coordinate Bethe ansatz, after Hans
Bethe, who first used the idea to find the eigenstates of the
Heisenberg spin chain~\cite{bethe}, and has since been applied to many
$1+1$ dimensional quantum and $2$ dimensional statistical mechanical
systems~\cite{various}.

Proceeding now with constructing the the eigenfunctions of~(\ref{se}), 
first consider a simpler model where $a_n=1$ and all other $a_l$
are equal to zero
\beq
\label{hamil}
       H= \sum_{j=1}^{N} (-i \partial_{x_j})^n + \sum_{j<k} 2c
       \delta(x_j - x_k)= T_n + V.
\eeq
The formulae for the eigenfunctions of this model are simpler and it
will be seen that the solution for the general model~(\ref{fullhamil})
can be easily constructed knowing the solution of~(\ref{hamil}).

Next, note that for coordinate values where no two $x_j$'s coincide,
$x_j \neq x_k$, the interaction term $V$ is zero and the eigenvalue
equation~(\ref{se}) reduces to 
\beq
\label{nonint}
      \sum_{j=1}^N (-i \partial_{x_j})^n \Psi = E \Psi.
\eeq
A ``basic'' solution of this equation is
\beq
\label{basic}
     e(x_1,..,x_N;k_1,...,k_N) = \exp i(\sum_{j=1}^N k_j x_j),
\eeq
where the $k_j$ are a set of quasimomentum parameters.  The energy and 
momentum of this state are
\beq
\label{ep}
        E= \sum_{j=1}^N k_j^n, \ \ \ \ \ \ \ \ \ \ \ \ \ P = \sum_{j=1}^N k_j.
\eeq

The effects of the interaction term $V$ may be rewritten as a set of
conditions which $\Psi$ must satisfy when crossing a boundary
$x_j=x_k$.  To obtain these conditions, integrate equation~(\ref{se})
across a boundary $x_j=x_k$:
\beq
\label{cond1}
     \lim_{\epsilon \rightarrow 0} \int_{-\epsilon}^{\epsilon}
     d(x_{k}-x_j) H \Psi = 
      \lim_{\epsilon \rightarrow 0} \int_{-\epsilon}^\epsilon
          d(x_k-x_j) E \Psi
\eeq
The RHS is zero, leaving only the kinetic and interaction terms.
Calculating the integral  
\beq
   \lim_{\epsilon \rightarrow 0} \int_{-\epsilon}^{\epsilon}
    d(x_k - x_j) (T_n + V) \Psi, 
\eeq
yields the equation
\beq
\label{bc}
   \frac{(-i)^n}{2^{n-1}} \sum_{l=0}^n {n \choose l} (1+(-1)^{n-l})
  ( ( \partial_{x_k}+\partial_{x_{j}})^l 
   ( \partial_{x_k}-\partial_{x_j})^{n-l-1} \Psi)|_{x_k = x_j}
   + c \Psi(x_k=x_j) = 0,
\eeq
where ${n \choose l}$ is the usual binomial coefficient. 
Thus, the orignal problem of solving~(\ref{se}) has been recast as
that of solving the noninteracting equation~(\ref{nonint}) subject to
the conditions~(\ref{bc}).

To simplify the task of solving~(\ref{nonint}) and~(\ref{bc}), note
that since this system describes bosons on a line, it is sufficient to
construct the eigenfunction only in an ordered sector,

\beq
                x_{Q_1}<x_{Q_2}<...<x_{Q_{N-1}}<x_{Q_N},
\eeq
where $Q$ is a permutation of $N$ elements. The value of the
wavefunction for other coordinate orderings can be obtained by
symmetric continuation.  For convenience, we set $Q=1$.

Next, consider the following superposition of the solutions~(\ref{basic})
\beq
\label{ba}
   \Psi= \sum_{P \in S_N} \phi(k_1,...,k_N;P) 
     \exp i (\sum_{j=1}^N k_{P_j} x_j),
\eeq
where $k_1,...,k_N$ are quasimomenta, $P$ is a permutation of $N$
elements, and $\phi(k_1,...,k_N;P)$ are $N!$ undetermined phase
functions of the quasimomenta and $P$.  The energy and momentum of
this state are given by~(\ref{ep}).

Now it will be determined how to choose the $\phi(P)$ so that the
boundary condition~(\ref{bc}) is satisfied across an intersection $x_j = 
x_{j+1}$.  Consider the partial sum
\beq
\label{pba}
  \tilde{ \Psi}=  \phi(P) \exp i (\sum_{j=1}^N k_{P_j} x_j) +
                    \phi(P') \exp i (\sum_{j=1}^N k_{P'_j} x_j),
\eeq
where $P'$ differs from $P$ by the exchange of $k_{P_j}$ and
$k_{P_{j+1}}$.  If this partial sum satisfies the boundary
conditions~(\ref{bc}), then so does the the entire sum~(\ref{ba})
since given a pair $j,j+1$, then $N!$ permutations may be split into
$\frac{N!}{2}$ pairs of the type in expression~(\ref{pba}).
Now requiring that the partial sum~(\ref{pba}) satisfies~(\ref{bc}),
one obtains a set of conditions on the phases $\phi(P)$
\beq
\label{pcc}
    \frac{-i}{2^{n-1}} \sum_{l=0}^n {n \choose l} (1+(-1)^{n-l}) 
    (k_{P_j} + k_{P_{j+1}})^l (k_{P_{j+1}} - k_{P_j})^{n-l-1}
                 (\phi(P) - \phi(P')) +
    c   (\phi(P) + \phi(P')) = 0.
\eeq
To write this in a more transparent form, define the polynomial
\beq
    P_n (k,k') =  \frac{1}{2^{n-1}} \sum_{l=0}^n {n \choose l} (1+(-1)^{n-l}) 
    (k' + k)^l (k' - k)^{n-l-1}.
\eeq
Note that this polynomial is of degree $n-1$ and odd under the
interchange $k$ and $ k'$.
Rewriting~(\ref{pcc}) in terms of the ratio of phases, one obtains
\beq
\label{sp}
   \frac{\phi(P')}{\phi(P)} =
   \frac{P_n(k_{P_j},k_{P_{j+1}}) + i c} {P_n(k_{P_j},k_{P_{j+1}}) - i
     c}
     \equiv \theta_n(k_{P_j},k_{P_{j+1}}).
\eeq
These equations have the interpretation as the phase
accrued when two particles scatter off one another. Note that the
scattering phase has the property $1/\theta_n(k,k') = \theta_n(k',k)$.

Lemma:A solution to the conditions~(\ref{sp}) is 
\beq
\label{soln}
     \phi(P) = \prod_{l>m, P_l<P_k} \theta_n(k_{P_l},k_{P_m}).
\eeq

Thus, the result for the eigenfunctions of~(\ref{hamil}) is
\beq
   \Psi= \sum_{P \in S_N} \phi(k_1,...,k_N;P) 
     \exp i (\sum_{j=1}^N k_{P_j} x_{Q_j}),
\eeq
with $\phi(P)$ chosen according to~(\ref{soln}), locally satisfies the
eigenvalue equation~(\ref{se}) for all natural $n$ and $N$.  The
eigenvalues of $H$ and $P$ are given by~(\ref{ep}).

Properties of the solution:

\begin{enumerate}

\item The wavefunction $\Psi$ is continuous for all $x_j$, and the
  first derivative $\partial_{x_j} \Psi(x)$ is discontinuous when any
  two $x$'s coincide

\item $\Psi (k_j=k_l) =0$ for all $j,k$, thus a Pauli-type exclusion
  principle holds for the quasimomenta parameters.  

\item Explicit formulae for the first few $\theta_n$ are
\beq
   n=1: \ \       \theta_1 = -1
\eeq
\beq
    n=2: \ \ \theta_2 = \frac{k - k' +i c}{k-k'-ic}
\eeq
\beq
    n=3: \ \ \theta_3 = \frac{\frac{3}{2}(k^2-k'^2) +  ic}
                             {\frac{3}{2}(k^2-k'^2) -  ic}
\eeq
\beq
    n=4: \ \ \theta_4 
    =\frac{\frac{1}{4}( 7 k^3 + 3k^2 k' - 3 k k'^2 - 7 k'^3) + ic}
          { \frac{1}{4} (7 k^3 + 3 k^2 k' - 3 k k'^2 - 7 k'^3) -  ic}
\eeq
\beq
    n=5 :  \ \ \theta_5 =
     \frac{ \frac{1}{8} (15 k^4 +10 k^3 k' - 10 k k'^3 -15 k'^4) + ic}
          { \frac{1}{8} (15 k^4 +10 k^3 k' - 10 k k'^3 -15 k'^4) - ic}
\eeq
The case $n=1$ corresponds to free fermions with a linear energy
spectrum.  
\end{enumerate}

Next, to construct the solution of the general
model~(\ref{fullhamil}), with arbitrary $a_l$, note that since the
differential equation~(\ref{se}) is linear, all polynomials
corresponding to different values of $l$ simply add.  So, stating the
main result:

{\bf Theorem}: the
eigenfunctions of~(\ref{fullhamil}) are of the form
\beq
   \Psi= \sum_{P \in S_N} \Phi(k_1,...,k_N;P) 
     \exp i (\sum_{j=1}^N k_{P_j} x_{Q_j}),
\eeq
where 
\beq
\Phi(P) = \prod_{l>m, P_l<P_k} \Theta_n(k_{P_l},k_{P_m}),
\eeq
and the two body scattering phase is
\beq
\Theta_n(k,k') = \frac{{\displaystyle \sum_{l=1}^{n} a_l P_l(k,k') + ic}}
                 {{ \displaystyle \sum_{l=1}^{n} a_l P_l(k,k') - ic}}.
\eeq
$\Theta_n$ obeys the properties $1/\Theta(k,k')= \Theta(k',k)$ and
$\Theta(k,k) = -1$.  The latter implies that the Pauli exclusion
principle for spectral parameters holds for the general model. The
eigenvalues of $H$ and $P$ are given by
\beq
     E= \sum_{j=1}^N \sum_{l=1}^n a_l (k_j)^l, \ \ \ \ \ \ \ \ \ \ 
     P = \sum_{j=1}^N k_j.
\eeq

Thus far, the boundary conditions on the partial differential
equation~(\ref{se}) acts have been left unspecified.  The next step
is to impose that the wavefunction $\Psi$ is periodic in a box of
length $L$, ie., require that
\beq
    \Psi(x_j+L) = \Psi(x_j) \ \ \ \ \ \ \ \ j=1,...,N.
\eeq
This condition imposes the constraint that the parameters
$k_1,...,k_N$ must satisfy a set of coupled equations
\beq
\label{bae}
      e^{i k_j L} =- \prod_{l=1}^{N} \Theta_n (k_j,k_l) \ \ \ \ \
    j=1,...,N.
\eeq
An interpretation of these equations is as the phase acquired by particle 
$j$ as it traverses around the box back to its starting point,
scattering off the other particles along the way.  

To understand the ground state, excitation spectrum, and other
physical characteristics of the hamiltonian~(\ref{hamil}), one needs
to analyze the solutions of the equations~(\ref{bae}).  The properties
of equations~(\ref{bae}) can differ essentially from those of
equations found for other $1+1$ dimensional quantum systems (e.g.
those of~\cite{various}).  In particular, for complex solutions, the
imaginary part of a solution can have a nontrivial functional
dependence on the real part. The properties of these complex solutions
need to be understood in order to study physical properties which
depend on the spectrum of eigenvalues.

The author thanks Minoru Takahashi for useful discussions and the
Japanese Society for the Promotion of Science for financial support.


\begin{thebibliography}{11}

\bibitem{korteweg} G.L. Lamb, {\it Elements of Soliton Theory} (Wiley
  Interscience, 1980), Ch. 6
\bibitem{weyl} H. Weyl, Ann. Phys. {\bf 59}, 101 (1919); Phys. Z. {\bf 
    22}, 473 (1021)
\bibitem{expand} A.A. Starobinsky, Sov. Atron. Lett. {\bf 9}, 302
  (1983); J.D. Barrow and A.C. Ottewill, J. Phys. A {\bf 16},2757
  (1983); M.B. Mijic, M.S. Morris, and W.-M. Suen, Phys. Rev. D{\bf
    34}, 2934 (1986)
\bibitem{renorm} K.S. Stelle, Phys. Rev D {\bf 16}, 953 (1977)
\bibitem{free} F.S. Fradkin and A.A. Tseytlin, Nucl. Phys. B{\bf 201}, 
  469 (1982)
\bibitem{nambu} T. Hamazaki and T. Kugo, Prog. Theo. Phys. {\bf 92},
  645 (1994)
\bibitem{schwinger} J. Barcelos-Neto and C.P. Natividae,
  Z. Phys. C{\bf 49}, 511 (1991)
\bibitem{skyrme} S. Dube and L. Marleau, Phys. Rev. D {\bf 41}, 1606 
  (1990); J. Neto, J. Phys. G {\bf 20}, 1527 (1994)
\bibitem{ss} J. Barcelos-Neto and N.R.F. Baga, Phys. Rev. D {\bf 39}, 494 
  (1989)
\bibitem{lieb} E.H. Lieb and W. Liniger, Phys. Rev. {\bf 130}, 1605-1616
  (1963); E. Lieb, ibid, 1617-1624
\bibitem{many} see the book {\it Quantum Inverse Scattering
    Method and Correlation Functions} (World Scientific Press, 1993) 
   by V.E.Korepin,
  N.M. Bogoliubov, and A.G. Izergin for a discussion and
  references.
\bibitem{bethe} H. Bethe, Z. Physik {\bf 71}, 225-226 (1931)
\bibitem{various} C.N. Yang and C.P. Yang, Phys. Rev. {\bf 150}, 321
  (1967); C.N. Yang, Phys. Rev. Lett. {\bf 19}, 1312 (1967); E.H.
  Lieb, Phys. Rev. {\bf 162}, 162 (1967); E.H. Lieb and F.Y. Wu, Phys.
  Rev. Lett. {\bf 20}, 1445 (1968); V.E. Korepin, Theo. Math. Phys.
  {\bf 41}, 953 (1979); H. Bergknoff and H.B. Thacker, Phys. Rev. D
  {\bf 19}, 3666 (1979)
 
\end{thebibliography}
\end{document}